\documentstyle[twoside,epsf,psfig]{article}

\catcode`\@=11
\long\def\@makefntext#1{
\protect\noindent \hbox to 3.2pt {\hskip-.9pt  
$^{{\eightrm\@thefnmark}}$\hfil}#1\hfill}		%CAN BE USED 

\def\thefootnote{\fnsymbol{footnote}}
\def\@makefnmark{\hbox to 0pt{$^{\@thefnmark}$\hss}}	%ORIGINAL 
	
\def\ps@myheadings{\let\@mkboth\@gobbletwo
\def\@oddhead{\hbox{}
\rightmark\hfil\eightrm\thepage}   
\def\@oddfoot{}\def\@evenhead{\eightrm\thepage\hfil
\leftmark\hbox{}}\def\@evenfoot{}
\def\sectionmark##1{}\def\subsectionmark##1{}}

\oddsidemargin=\evensidemargin
\renewcommand{\thefootnote}{\fnsymbol{footnote}}

\newcounter{sectionc}\newcounter{subsectionc}\newcounter{subsubsectionc}
\renewcommand{\section}[1] {\vspace{12pt}\addtocounter{sectionc}{1} 
\setcounter{subsectionc}{0}\setcounter{subsubsectionc}{0}\noindent 
	{\tenbf\thesectionc. #1}\par\vspace{5pt}}
\renewcommand{\subsection}[1] {\vspace{12pt}\addtocounter{subsectionc}{1} 
\setcounter{subsubsectionc}{0}\noindent 
{\bf\thesectionc.\thesubsectionc. {\kern1pt \bfit #1}}\par\vspace{5pt}}
\renewcommand{\subsubsection}[1] {\vspace{12pt}\addtocounter{subsubsectionc}{1}
	\noindent{\tenrm\thesectionc.\thesubsectionc.\thesubsubsectionc.
	{\kern1pt \tenit #1}}\par\vspace{5pt}}
\newcommand{\nonumsection}[1] {\vspace{12pt}\noindent{\tenbf #1}
	\par\vspace{5pt}}

\newcounter{appendixc}
\newcounter{subappendixc}[appendixc]
\newcounter{subsubappendixc}[subappendixc]
\renewcommand{\thesubappendixc}{\Alph{appendixc}.\arabic{subappendixc}}
\renewcommand{\thesubsubappendixc}
	{\Alph{appendixc}.\arabic{subappendixc}.\arabic{subsubappendixc}}

\renewcommand{\appendix}[1] {\vspace{12pt}
        \refstepcounter{appendixc}
        \setcounter{figure}{0}
        \setcounter{table}{0}
        \setcounter{lemma}{0}
        \setcounter{theorem}{0}
        \setcounter{corollary}{0}
        \setcounter{definition}{0}
        \setcounter{equation}{0}
        \renewcommand{\thefigure}{\Alph{appendixc}.\arabic{figure}}
        \renewcommand{\thetable}{\Alph{appendixc}.\arabic{table}}
        \renewcommand{\theappendixc}{\Alph{appendixc}}
        \renewcommand{\thelemma}{\Alph{appendixc}.\arabic{lemma}}
        \renewcommand{\thetheorem}{\Alph{appendixc}.\arabic{theorem}}
        \renewcommand{\thedefinition}{\Alph{appendixc}.\arabic{definition}}
        \renewcommand{\thecorollary}{\Alph{appendixc}.\arabic{corollary}}
        \renewcommand{\theequation}{\Alph{appendixc}.\arabic{equation}}
        \noindent{\tenbf Appendix \theappendixc #1}\par\vspace{5pt}}
\newcommand{\subappendix}[1] {\vspace{12pt}
        \refstepcounter{subappendixc}
        \noindent{\bf Appendix \thesubappendixc. {\kern1pt \bfit #1}}
	\par\vspace{5pt}}
\newcommand{\subsubappendix}[1] {\vspace{12pt}
        \refstepcounter{subsubappendixc}
        \noindent{\rm Appendix \thesubsubappendixc. {\kern1pt \tenit #1}}
	\par\vspace{5pt}}

\newcommand{\textlineskip}{\baselineskip=13pt}
\newcommand{\smalllineskip}{\baselineskip=10pt}

\newcommand{\copyrightheading}[1]
	{\vspace*{-2.5cm}\smalllineskip{\flushleft
	{\footnotesize International Journal of Modern Physics D, #1}\\
	{\footnotesize \copyright\kern2pt World Scientific Publishing
	 Company}\\
	 }}

\newcommand{\publisher}[2]{{\begin{center}\footnotesize\smalllineskip 
	Received #1\\
	Revised #2
	\end{center}
	}}

\def\abstracts#1#2#3{{
	\centering{\begin{minipage}{4.5in}\footnotesize\baselineskip=10pt
	\parindent=0pt #1\par 
	\parindent=15pt #2\par
	\parindent=15pt #3
	\end{minipage}}\par}} 

\renewenvironment{thebibliography}[1]
        {\frenchspacing
	 \ninerm\baselineskip=11pt
         \begin{list}{\arabic{enumi}.}
        {\usecounter{enumi}\setlength{\parsep}{0pt}     
	 \setlength{\leftmargin 12.7pt}{\rightmargin 0pt}%FOR 1--9 ITEMS
         \setlength{\itemsep}{0pt} \settowidth
	{\labelwidth}{#1.}\sloppy}}{\end{list}}

\newcounter{itemlistc}
\newcounter{romanlistc}
\newcounter{alphlistc}
\newcounter{arabiclistc}

\newcommand{\fcaption}[1]{
        \refstepcounter{figure}
        \setbox\@tempboxa = \hbox{\footnotesize Fig.~\thefigure. #1}
        \ifdim \wd\@tempboxa > 5in
           {\begin{center}
        \parbox{5in}{\footnotesize\smalllineskip Fig.~\thefigure. #1}
            \end{center}}
        \else
             {\begin{center}
             {\footnotesize Fig.~\thefigure. #1}
              \end{center}}
        \fi}

\newcommand{\tcaption}[1]{
        \refstepcounter{table}
        \setbox\@tempboxa = \hbox{\footnotesize Table~\thetable. #1}
        \ifdim \wd\@tempboxa > 5in
           {\begin{center}
        \parbox{5in}{\footnotesize\smalllineskip Table~\thetable. #1}
            \end{center}}
        \else
             {\begin{center}
             {\footnotesize Table~\thetable. #1}
              \end{center}}
        \fi}

\def\@citex[#1]#2{\if@filesw\immediate\write\@auxout
	{\string\citation{#2}}\fi
\def\@citea{}\@cite{\@for\@citeb:=#2\do
	{\@citea\def\@citea{,}\@ifundefined
	{b@\@citeb}{{\bf ?}\@warning
	{Citation `\@citeb' on page \thepage \space undefined}}
	{\csname b@\@citeb\endcsname}}}{#1}}

\newif\if@cghi
\def\cite{\@cghitrue\@ifnextchar [{\@tempswatrue
	\@citex}{\@tempswafalse\@citex[]}}
\def\citelow{\@cghifalse\@ifnextchar [{\@tempswatrue
	\@citex}{\@tempswafalse\@citex[]}}
\def\@cite#1#2{{$\null^{#1}$\if@tempswa\typeout
	{IJCGA warning: optional citation argument 
	ignored: `#2'} \fi}}

\def\pmb#1{\setbox0=\hbox{#1}
	\kern-.025em\copy0\kern-\wd0
	\kern.05em\copy0\kern-\wd0
	\kern-.025em\raise.0433em\box0}

\def\fnt#1#2{\footnotetext{\kern-.3em
	{$^{\mbox{\scriptsize #1}}$}{#2}}}

\def\fpage#1{\begingroup
\voffset=.3in
\thispagestyle{empty}\begin{table}[b]\centerline{\footnotesize #1}
	\end{table}\endgroup}

\def\runninghead#1#2{\pagestyle{myheadings}
\markboth{{\protect\footnotesize\it{\quad #1}}\hfill}
{\hfill{\protect\footnotesize\it{#2\quad}}}}
\font\tenrm=cmr10
\font\tenit=cmti10 
\font\tenbf=cmbx10
\font\bfit=cmbxti10 at 10pt
\font\ninerm=cmr9

\font\eightrm=cmr8

\def\qed{\hbox{${\vcenter{\vbox{	          %HOLLOW SQUARE
   \hrule height 0.4pt\hbox{\vrule width 0.4pt height 6pt
   \kern5pt\vrule width 0.4pt}\hrule height 0.4pt}}}$}}

\renewcommand{\thefootnote}{\fnsymbol{footnote}}  %USE SYMBOLIC FOOTNOTE

\begin{document}
\setlength{\textheight}{7.7truein}    %FOR 2ND PAGE ONWARDS

\runninghead{Disagreements of Experiments with Theories on Gravitation and Cosmology $\ldots$} {Disagreements of Experiments with Theories on Gravitation and Cosmology $\ldots$}

\normalsize\textlineskip
\thispagestyle{empty}
\setcounter{page}{1}

\copyrightheading{}		%{Vol.~0, No.~0 (1999) 000--000}

\vspace*{0.88truein}

\fpage{1}
\centerline{\bf FUNDAMENTAL DISAGREEMENTS OF EXPERIMENTS}
\vspace*{0.035truein}
\centerline{\bf WITH THEORIES ON}
\vspace*{0.035truein} 
\centerline{\bf GRAVITATION AND COSMOLOGY}
\vspace*{0.37truein}
\centerline{\footnotesize RAFAEL A. VERA\footnote{Galvarino 482 
Concepci¢n. Chile rvera@udec.cl.}}

\baselineskip=12pt

\centerline{\footnotesize\it Departamento de F¡sica, Universidad 
de Concepci¢n, Concepci¢n. Chile}

\baselineskip=10pt
 \centerline{\footnotesize\it Concepci¢n, Correo 3, Chile}

\vspace*{0.225truein}
\publisher{(received date)}{(revised date)}

\vspace*{0.21truein}
\abstracts{
From the equivalence principle (EP) and experiments on gravitational (G) 
time dilation (GTD) it is proved that the standards of observers located 
in different ``distances'' from the earth are physically different with 
respect to each other. Thus the current mathematical relationships 
between their measurements are physically inhomogeneous. This has caused 
fundamental errors in gravitation, cosmology and astrophysics. The true 
transformations between basic parameters of bodies located in different 
field positions, derived from just experimental facts, are used to test 
fundamental hypotheses in current literature. G fields do not exchange 
energy with bodies and radiation, but just momentum. The G energy comes 
the bodies. The average relative distances and cosmological redshifts in 
the universe cannot change after universe expansion because the average 
increase of distances (G potentials) would change the sizes of particles 
in identical proportion. Locally, atoms must be evolving, indefinitely, in 
closed cycles between states of gas and ``linear black hole''. The last 
ones, after recovering energy, must explode thus regenerating gas. Most 
of the universe must be in state of black galaxy cooled down by linear 
black holes. They must account for the CMBR. The new scenarios are also 
explained by using particle models consistent with the EP.}{}{}
%\vspace*{10pt}
%\keywords{The contents of the keywords}

\vspace*{1pt}\textlineskip	
\section{A Source of Fundamental Inexactitudes in Physics}
\vspace*{-0.5pt}
\subsection{The tacit postulate of physics}

\noindent
The conventional tests for gravitational theories and
the Einstein's theory on gravitation were conceived long ago from direct
relationships between quantities measured by observers at rest in different
positions of G fields. In them, it is assumed that such observers have
reference standards that are physically the same compared to each other. For
this reason, the symbols for quantities such as the rest mass of a body, or
the frequency of a clock, are not position dependent. Then all of them, general
relativity (GR) and the G tests are based on the same ``tacit postulate'' on
the absolute invariability of the bodies after a change of position in a G
field.
 
\setcounter{footnote}{0}
\renewcommand{\thefootnote}{\alph{footnote}}

\subsection{``The top of the iceberg'' (previous discussion)}

On the other hand, such postulate is not consistent with the true 
G time dilation (GTD) experiments in which the readings of clocks positioned 
in different distances from the earth, for relatively long periods, have 
been compared to each other. From such experiments, it is clear that some
fundamental physical changes have occurred to the clocks during the whole
period in which they have been in a G potential different to the original
one.

It is important to observe that such experiments have measured time
intervals that are entirely independent on what may happen to some 
secondary photons used in some of these experiments.

For example, in the experiments of Hafele-Kerting (1972) no photons were 
used\cite{HafeleK}. In them, the readings of the clocks, before and after 
trips of 48 hours with an average height of about $9$ km, were compared, 
locally, with a reference clock in the earth surface. The
GTD components, after correction for kinetic time dilation, were clearly
positive. Then there is not doubt in that, during 48 hours, the flying
clocks had been running with a higher frequency compared to the one in 
the land.

The Global Positioning System (GPS) is also equivalent to a
permanent GTD experiment. This is because the orbiting clocks were
``specially designed'' with a pre-calculated frequency, in the earth
surface, that took into account both the kinetic and the gravitational time
dilation. ``Their original frequencies were significantly lower than of the 
one of an earth
standard clock''. So that the final frequency, in orbit, is equal to the one
of an earth standard clock. Then, there is no doubt in that the orbiting
clocks are running with a higher frequency with respect to the ``original''
clocks in the earth surface. This is obviously due to the positive
contribution of GTD.

The same conclusion comes out from the GTD experiment made up
by Vessot and Levine (1979)\cite{Vessot}. A highly accurate hydrogen-maser
clock was launched on a scout rocket up to an elevation of about $10,000$ km
. Well-defined time-intervals of the flying clock were compared, ``via radio
signals'', with the ones of a clock on the ground. The results of such
experiment, corrected for velocity differences, confirmed the real
existence of GTD to within $2$ parts in $10^{4}$.

The same as in the above cases, {\it the results of the
Vessot-Lavine experiments, corrected for differences of velocity, did not 
depend on anything that may have occurred
to the radio signals during their trips between the clocks}. The time
intervals are differences of times of consecutive radio signals in which the
time of flight of the photons is canceled out. Such time intervals are
independent on the frequency of such photons.

Then, from the results of GTD experiments it is concluded that:

\begin{enumerate}
 
  \item - The frequency of the standard clocks of observers located at 
rest in different distances from the earth depend on their distances from
the earth center. According to the equivalence principle (EP)\cite{Misner}, 
the local physical laws remain unchanged only if other parameters of atoms
and clocks change in the same proportion. This means that ``their reference 
standards that are physically different with respect to each other''.

  \item - {\it The direct relations between frequencies,
energies and masses measured by two observers located at rest in different G
potentials are not strictly homogeneous}. They have not a well-defined
physical meaning because their reference standards are physically different
with respect to each other.
\end{enumerate}

The main purpose of this article is to prove that this kind of
error has been a source of a chain of other fundamental errors in the
current literature on gravitation, cosmology, astrophysics and astronomy.
Such errors have produced important misinterpretations of a relatively large
number of gravitational and celestial phenomena.

\section{The Nonlocal Formalism Fixed by Experimental Facts}

\noindent
Since the experiments prove that bodies are not are not physically 
invariable, after a change of distance with respect to the G field source, 
then the current formalisms that assume the physical invariability of the 
bodies must be not used here. 

For a self-consistent description of the real changes that have occurred to
a NL clock, after a change of position in a G field, it is essential that
the observer does not change of position in such field. This is equivalent
to use a more ``strictly invariable'' (SI) reference clock that does not
change of position and velocity with respect to he G field sources. This one
fixes a strictly invariable (flat) theoretical reference frame that in
principle can be used to describe the real properties of the NL bodies and 
of the NL space in a G field.

To relate quantities measured by observers located at rest in different 
positions of a static G field, all of them must be previously 
``transformed'' to some common SI reference frame based on some
well- defined standard body located at rest in some well-defined position of
such field. 

The basic relationships, or factors, used to correct the quantities for 
differences of G potentials between objects and observers, are called here 
``gravitational transformations''.

\subsection{Conventions}
\noindent 
Due to the physical changes occurring to the bodies, after a change 
of G potential, it is necessary to make a distinction
between the cases in which the object and observer are located in the same
potential and the ones in which they are located in clearly different ones. 
Here, to typify such cases, the words {\it local and nonlocal} 
(NL), respectively,  have been used\footnote{These words 
are used for just the cited purposes. They are not
necessarily related to other definitions done for different purposes in the
current literature. Practical limits between the two cases can be defined
later one for each specific case. For the moment, let us assume that a local
observer is infinitely close to his local objects so that the differences of
G potential between them can be neglected.}.

Normally, the results of the GTD experiments have
been already corrected for kinetic time dilation, due to earth rotation or
other eventual movements of the clocks. This has been normally done after
Lorenz transformations. Then such results correspond to the idealized case
in which {\it bodies and clocks would be at rest in a non-rotating earth}.
For this reason, and for simplicity, below, all of the bodies (objects and
observers) have been assumed to be at rest in a strictly static central
field. In such idealized case, it is assumed that some {\it SI observer} and
his local standard clock, called $A$, stays at rest in the ground, in some
fixed radius ($a$) from the earth center. The NL objects are at rest in
general positions ($r$) from such center.

According to the results of the GTD experiments, the ratio between the 
frequency of the NL clock and the one of the local clock depends both on $r$
and $a$. {\it Since the observer is in a fixed position $a$ of the G field 
then this one is stated by means of a subscript}. Such position fixes the 
well-defined frequency of the reference clock, which in turn indirectly fixes 
the precise physical unit system of such quantity.

For example, the generic symbol for the frequency of a NL clock
``at rest'' at $r$, with respect to a SI observer at $a$, is: $\nu_{a}(0,r)$
. The quantities in the parenthesis are the velocity ($V=0$) and position of
the clock with respect to the earth center.

According to the traditional convention, two observers at rest
in different distances from the earth center, like $a$ and $r$, assign
identical ``numerical values'' to the frequencies of their ``local''
standard clocks, say $1Hz$. 

On the other hand, according to GTD experiments,
such clocks run with different frequencies with respect to each other, i.e.,
they are not physically identical with respect to each other. Thus, to make
a clear identification of the position of the local clock, the
better-defined symbols like $\nu_ {a}(0,a)$ and $\nu_ {r}(0,r)$ must be used
instead of the invariable numerical values used in the current literature.
In this way the position of the reference standard is clearly stated so as
to prevent physical inhomogeneities in the equations, as shown below.

In more general cases, when a body is ``moving'' with respect
to the observer, according to the Lorenz transformations, the parameters of
the NL body, with respect to the fixed observer, must also depend on the
velocity of the body with respect to the observer.  For this reason, the
symbol for the NL mass of a body at $r$, falling with the velocity $V$, with
respect to the SI observer at $a$, is $m_{a}(V,r)$.

The formalism fixed by the experimental facts
corresponds to a plain generalization of the one used in special
relativity. Then a reasonable name for it is ``NL relativity''.

For the present purposes and for the specific cases used here,
it is not necessary to use explicit four-dimensional relationships. 
Below it is shown
that there are advantages of treating each parameter of a body as a
separated member of a multiple equation. In this way it is more obvious what
happens to each of them, after a common change of G potential.

\section{The Nonlocal Form of the Equivalence Principle (NL EP)}
\noindent
The EP has two alternatives of solutions: The conventional one, that 
matter is invariable after a change of position with respect to the field 
source, and the opposite one, that ``all of the parameters of any part of the 
system change in the same proportion''. The last alternative is the single one
consistent with the true results of the GTD experiments. 

\subsection {Application of the EP to the GTD experiments}

\noindent
Assume two measuring systems, $A$ and $B$, located
at rest in the distances $a$ and $r$ from the earth center, respectively.
Let us call $\nu_{r}^{atom}(0,r)$ to the ``local'' frequency of some
particular spectrum line of some atom at $B$. This is a ratio between the 
frequency of such atom and the frequency of some standard clock located in 
just the same G potential of the atom.

Since this ratio is dimensionless, then it has the same value
even when the ``two'' measurements are made up by a ``nonlocal'' observer 
$A$, which is positioned in a another G potential.

\noindent 
\begin{equation}
\nu_{r}^{atom}(0,r)= \frac{\nu_{a}^{atom}(0,r)} {\nu_{a}^{std}(0,r)}=
Constant \,.  \label{1}
\end{equation}

In which $\nu_{a}^{atom}(0,r)$ and $\nu_{a}^{st}(0,r)$ are the
``NL'' frequencies of the atom and of the standard clock at $r$, with
respect to the observer at $a$.

Assume now that the system $B$ is moved from $r$ up to a new
rest at $r+dr$. According to the EP, the observer at $B$ cannot detect the
changes of frequency of ``his'' local clock. Then the derivative of the
logarithm of Eq.~(\ref{1}) must be zero\footnote{Since
the constant values of velocity and position are explicitly stated, then the
single variable is $r$. Thus it is unnecessary to use symbols of partial
derivatives}.

\noindent 
\begin{equation}
\frac{d\nu_{r}^{atom}(0,r)}{\nu_{r}^{atom}(0,r)}= \frac{d\nu_{a}^{atom}(0,r)%
}{\nu_{a}^{atom}(0,r)}- \frac{d\nu_{a}^{std}(0,r)}{\nu_{a}^{std}(0,r)}=0\,.
\label{2}
\end{equation}

On the other hand, according to the positive results of the GTD
experiments, each of the two terms of last member are not zero:

\noindent 
\begin{equation}
\frac{d\nu_{a}^{atom}(0,r)}{\nu_{a}^{atom}(0,r)}= \frac{d\nu_{a}^{std}(0,r)}{
\nu_{a}^{std}(0,r)}= d\phi\neq 0\,.  \label{3}
\end{equation}

in which $d\phi$ is the proportion observed from GTD. Since Eq. (\ref{3}) is 
dimensionless, this proportion must be
the same for all of observers at rest with respect to each other, regardless
of their positions in the field. This means that, {\it all of the
eigen-frequencies, of every part of the system $B$, have changed in just the
same proportion as the frequency of the clock at $B$, with respect to the
observer $A$ that has not changed of G potential}.

On the other hand, from to EP, the ``local'' parameters of any
well-defined part of the system are related to each other by particular
constants that do not change after any common change of G potential. This
holds, for example, for any frequency, mass-energy, length and wavelength of
any well-defined part of the same system. This can be expressed by a generic
expression like:

\noindent 
\begin{equation}
k^{\nu }\nu_{r}(0,r)=k^{m}m_{r}(0,r)=k^{\lambda }\lambda_{r}(0,r)={k^L}%
L_{r}(0,r)\,.  \label{4}
\end{equation}

in which $k^{\nu }$, $k^{m}$, $k^{\lambda }$, and $k^{L}$, are
particular constants for each case. Thus relationships similar to Eq.~(\ref
{3}) must hold for each of the parameters in Eq.~(\ref{4}). All of them must
change in the same proportion given by:

\noindent 
\begin{equation}
\frac{d\nu_{a}(0,r)}{\nu_ {a}(0,r)}=\frac{dm_{a}(0,r)}{m_{a}(0,r)}=\frac{
d\lambda_{a}(0,r)}{\lambda_{a}(0,r)}=\frac{dL_{a}(0,r)}{L_{a}(0,r)}
=d\phi\,.  \label{5}
\end{equation}

Then, the positive results of the GTD experiments can be
consistent with the EP only if: ``{\it after a change of G potential of a
measuring system, the basic parameters of all of its well-defined parts
change linearly, in just the same proportion as the frequencies of the
clocks in such system''}. This may be called the NL form of the EP (NL EP).

According to special relativity, a similar fact holds for the
relative changes occurring to the vectors of every well-defined part of a
system that changes of velocity with respect to the SI observer. Then, to
agree with all of the experimental facts it is necessary to admit that {\it
all of the well-defined parts of the local system obey the same inertial and
gravitational laws. All of their vectors must change in the same way and in
an identical proportion after any common change of velocity or position with
respect to the G source}. This means that the corresponding relationships
must be strictly ``linear''.

Notice that this is the single solution that can account for
the results of all of the most exact experiments made up in local and ``NL
conditions'' in G fields. If the inertial and/or gravitational laws of any
well-defined (measurable) part of a system were different from another part
of it, then such differences could be detected from local measurements made
up after changes of velocity and/or G potentials of the measuring system.
Such positive detections would violate the EP, which has never occurred.

Indeed, the NL EP is a principle on the common nature of every
well-defined part (or particle) of any particular system. 
More specifically, in that {\it all of the parts of the system must obey 
the same inertial and gravitational laws}, and that such laws must be 
strictly linear ones.

Consequently, the NL EP must also hold for any quantum of
radiation stationary state that may exist in a wave cavity of the same
system. This is because its well-defined eigen-values of frequency and
wavelength are related to the parameters of the wave cavity by particular
constants that do not change after changes of G potential. 
Thus, particle models made up of photons in
stationary states can emulate uncharged particles\footnote{This
model is used below to account for the non- conventional results of the
experiments. A similar model was successfully used in previous works to get
theoretical properties of ordinary bodies and of their G fields on the base
of general properties of radiation\cite{Vera81b}$^,$\cite{Vera97}.}.

\section{The G Transformations Fixed by GTD Experiments}
\subsection{The GTD experiments}

\noindent 
Assume a GTD experiment in which the observer $A$ has two local 
clocks of frequency $\nu_ {a}(0,r)$. One of them is raised from a 
radius $r=a$ up to a new rest at $r=r+dr$, in which 
$dr$ is very small compared with $r$.  According to the results of the such 
experiment ``the proportional change of frequency the NL clock'', 
compared with its initial frequency, is defined by the first member of the 
next equation. The next members, give the results of such experiments in 
terms of different measurement alternatives:

\noindent 
\begin{equation}
\frac{\nu_{a}(0,r+dr)-\nu_{a}(0,r)}{\nu_{a}(0,r)}\cong \frac{g_{a}(r)dr_{a}}{
c_{a}^{2}}\cong \frac{G_{a}M_{a}}{r_{a}}- \frac{G_{a}M_{a}}{r_{a}+dr_{a}}=
\frac{dE_{a}(r)}{m_{a}(0,r)}=d\phi (r)\,.  \label{6}
\end{equation}

The second member of (\ref{6}) is given in terms of the average
acceleration of gravity within $r$ and $r+dr$. The third one is done in
terms of the initial and final positions of the NL clock with respect to the
G field source\footnote{The masses and energies are expressed
here in joules. Thus the numerical value of $G$ is $G[mks].c^{-4}$.}

The fourth member is given in terms of the exact proportion of the energy 
given up to the clock of mass $m_{a}(0,r)$ to raise it, from $r=a$ up to 
$a+dr$. According to current experiments this proportion is the same
for any test body. It is also identical to the proportion of energy released
during the stop, after a free fall from $r$. This member has 
been preferred below when more exact and better defined values are required.
The last member is the symbol for the dimensionless change of NL potential 
defined by this equation.

Since this is a dimensionless equation, each member of Eq.~(\ref
{6}) is independent on the position of the SI observer. 
This property puts into relief that this fraction is a
measure of a real (absolute) kind of change that has occurred 
to the body, after a change of distance from the G field source.

\subsection{The G transformations fixed by the by the NL EP}

\noindent
The results of the GTD experiments, given in Eq.~(\ref
{6}), must fit with the conditions imposed by the EP, which is given in 
Eq.~(\ref{5}). Then, the different parameters of any well defined part 
of the system must change in the same proportion, i.e., 
according to:

\noindent 
\begin{equation}
\frac{d\nu_{a}(0,r)}{\nu_ {a}(0,r)}=\frac{dm_{a}(0,r)}{m_{a}(0,r)} =\frac{%
d\lambda_{a}(0,r)}{\lambda_{a}(0,r)} =\frac{dL_{a}(0,r)}{L_{a}(0,r)} =%
\frac{dE_{a}(r)}{m_{a}(0,r)}=d\phi (r)\,.  \label{7}
\end{equation}

In the case of any stationary photon in the NL clock, or in a wave cavity, 
from the wave properties of radiation, its average NL speed with 
respect to the observer is related to the average values of its NL 
frequency and of its NL wavelength by:

\noindent 
\begin{equation}
c_{a}(r)=\nu_{a}(0,r)\lambda _{a}(0,r)\,.  \label{8}
\end{equation}

From ~(\ref{7}) and (\ref{8}), the proportional changes of the
basic variables of the stationary photon  turns out to be related to each 
other by:

\noindent 
\begin{equation}
\frac{d\nu_{a}(0,r)}{\nu_{a}(0,r)} =\frac{d\lambda_{a}(0,r)}{\lambda
_{a}(0,r)} =\frac{1}{2}\frac{dc_{a}(r)}{c_{a}(r)}\,.  \label{9}
\end{equation}

The integration of (\ref{7}) and (\ref{9}), for a body that was
moved from the position $a$ of the observer up to a general position $r$,
gives a net G transformation factor:

\noindent 
\begin{equation}
f_{a}(r)=\frac{\nu_{a}(0,r)}{\nu_{a}(0,a)} =\frac{m_{a}(0,r)}{m_{a}(0,a)} =%
\frac{\lambda_{a}(0,r)}{\lambda_{a}(0,a)} =\sqrt{\frac{c_{a}(r)}{c_{a}(a)}}
=e^{\Delta \phi _{a}(r)}\cong 1 +\frac{\Delta E_{a}(r)}{m_{a}(0,a)}\,.
\label{10}
\end{equation}

From the EP, the first four members hold for any well-defined
frequency, mass, length or wavelength, of any well-defined part of the same
NL system at rest with respect to the observer.

The fifth member gives the explicit values of the
transformation factors in terms of the NL speed of light in each place. This
member reveals that ``the G field is a gradient of the NL refraction index
of the space''.

Notice that {\it the NL speed of light is the one that fixes
the eigen-values of the NL frequencies and NL wavelengths of any
well-defined particle at rest in the field}\footnote{This
position dependent speed of light is not in conflict with 
special relativity, because in any
local limit, when $r\rightarrow a$, the numerical value of 
$c_{r}(a)\rightarrow c$, regardless of the value of $a$.}.

The sixth member is expressed in terms of the difference of the
G potential that exists between the object and the observer. This one is a
function of the positions of the body and of the observer, according to:

\noindent 
\begin{equation}
\Delta \phi (r)\cong \frac{GM_{a}}{a} -\frac{GM_{a}}{r}\,.  \label{11}
\end{equation}

The last member of Eq.~(\ref{10}) is expressed in terms of the
fraction of mass-energy given up to the body, to raise it from $a$ up to $r$.

From Eq.~(\ref{10}) it is obvious that the parameters of a NL
object at rest in the field, with respect to the observer, are point
functions that depend only on the positions of the NL object and of the
observer's reference standard. Something similar holds for the NL speed of
light.

This equation shows that in regions of lower NL speed of light
there is a general {\it G contraction} of the rest-values of the basic
parameters of bodies and radiation in stationary state. They depend on the
square root of the NL refraction index of the space, with respect to the one
of the SI observer.

\subsection{Consistency with other gravitational experiments}

\noindent
Eq.~(\ref{10}) is obviously consistent with the
EP because the value of $f_{r}(r)$ is $1$ for any value of $r$. This value
is independent on the velocity and position of the local system with respect
to the rest of the universe. This means a full agreement with the most exact
``local'' experiments.

The NL EP and Eq. (\ref{10}), are clearly verified from the
fact that the percentages of G redshifts are independent on the actual
frequencies. This means that ``the frequencies of the clocks and of all of
the spectrum lines of any atom turn out to be shifted in just the same
proportion after identical changes of position with respect to the G field
sources.''

The consistency of Eq. (\ref{10}) with the 
conventional G tests, by using the present formalism, has
been verified by Vera (1981)in a previous 
work\cite{Vera81b}. In it the Eq.~(\ref{10}) was
derived in an entirely different way. This was done step by step from
theoretical properties of a particle model made up of photons in stationary
state. This equation was applied to the conventional experiments used for
testing G theories. The results are consistent with the experimental
values.
Indeed the G transformations cannot be in contradiction with the experiments 
because they are just the results of the experimental data. For this reason
they can be directly used, to find answers for some fundamental questions 
in gravitation and cosmology, independently on any theoretical approach.

\section{Can photons exchange energy with static G fields?}

\noindent In ordinary physics it is currently assumed that the 
redshift of light is due to some exchange of energy between the 
photons and the G field, which would occur during the 
trip of the photons. 

It is simple to verify that such hypothesis is a consequence of assuming 
that the clocks of all of the observers run with the same frequency 
regardless on their positions in the field. Since such clocks have 
different frequencies, then the opposite alternative is expected to be 
the right one. This can be demonstrated from several 
different ways, as follows:

\subsection{Demonstration from combining experiments on
GTD and G redshift}

\noindent
According to the NL EP, and Eq. (\ref{3}), the eigen-frequencies of all 
of the parts of the NL system at $r$ are shifted in just the same 
proportion with respect to the observer at $a$. In particular, this holds 
for all of them: for the 
eigen-frequency of the NL clock, called $\nu_{a}(0,r)^{clock}$, 
for the eigen-frequency of
any particular NL atom, called $\nu_{a}(0,r)^{atom}$, and for the 
NL frequency of the photon emitted by such atom, called 
$\nu_{a}(r)^{photon}$.

\noindent 
\begin{equation}
\Delta{\phi}=\frac{\Delta{\nu_{a}(0,r)^{clock}}}{\nu_{a}(0,a)^{clock}}=
\frac{\Delta{\nu_{a}(0,r)^{atom}}}{\nu_{a}(0,a)^{atom}} =\frac{\Delta{
\nu_{a}(r)^{photon}}}{\nu_{a}(0,a)^{atom}} \,.  \label{12}
\end{equation}

On the other hand, according to the results of the experiments
on GTD and GRS, the frequency of the photons at the end of the trip, 
called $\nu_{a}(0,a)^{atom}$, are shifted in just the same proportion 
as the clock, i.e.,

\noindent 
\begin{equation}
\Delta{\phi}=\frac{\Delta{\nu_{a}(a)^{photon}}} {\nu_{a}(0,a)^{atom}}\,.
\label{13}
\end{equation}

Thus, from ~(\ref{12}) and (\ref{13}):

\noindent 
\begin{equation}
\nu_{a}(r)_{photon} =\nu_{a}(a)_{photon} \ ,.  \label{14}
\end{equation}

This means that:

\begin{itemize}
\item  During the trip of a photon, in a G field, its
frequency with respect to a clock in a fixed position of such field, 
remains constant (NL frequency conservation of free photons.).

\item  Photons do not exchange energy with static G fields.
\end{itemize}

\subsection{Demonstration from wave properties of light}

\noindent
According to {\it wave-continuity} the waves of a wavetrain cannot be 
suddenly lost or created by a strictly static conservative field. Then the 
number of them cannot change during its trip throughout the gradient of 
the NL refraction index of the G field revealed by Eq. (\ref{10}). The 
same holds for the number of waves (or wavelets) per unit of SI time 
crossing any plane static with respect to the observer. In other terms, 
the NL frequency of light, with respect to any clock of constant frequency, 
must also remain constant.

\subsection{Demonstrations from simultaneous fit with other G tests.}

\noindent
In a previous work\cite{Vera81b}, Vera derived Eq. (\ref{10}) from a 
theoretical approach and demonstrated proved that it accounts for
the conventional G tests provided that {\it the NL frequency of the 
radiation remains constant with respect to an observer in a fixed potential}.

\subsection{Demonstration from refraction properties the space}

\noindent
From Eq.~(\ref{10}), the deviation of light in a G
field is just a refraction phenomenon. {\it According to refraction laws,
the photon's color (frequency) does not change during refraction produced by
a static dielectric.} Then light does not exchange energy with the G field
because photons do not exchange energy with static dielectrics. They
exchange just {\it momentum}.

\section{Is it True that G Fields Exchange Energy with Bodies?}

\noindent
Einstein conceived the field equation of general relativity on the base 
that, during G work, {\it the G field gives up energy to the 
body}\cite{Einstein}. Let us find whether or not such hypothesis is 
consistent with the experimental facts described by strictly
homogenous transformations.

\subsection{Demonstration from G transformations}

\noindent
Assume that the observer at $a$ throws upwards a body with an energy 
$\Delta E_{a}(a)$. The body becomes at rest in the position of its 
maximum radius $r$. From the second and the fifth members of
Eq.~(\ref{7}), the mass-energy of the body at rest at $r$ with respect to
the observer at $a$ has increased by the amount:

\noindent 
\begin{equation}
\Delta m_{a}(0,r)=\Delta E_{a}(a)\,.  \label{15}
\end{equation}

This means that {\it all of the energy given up to the body is
transformed into an additional mass-energy of the same body}. Then such
energy is not spread in the field. It is just a fraction of the mass-energy
of the same body, with respect to such observer. No energy at all has been
transferred to the G field.

Vice versa, for the free fall form $r$ up to $a$, from Eq.~(\ref
{10}), the initial (NL) rest mass of the test body at $r$, with respect to
the observer at $a$, is:

\noindent 
\begin{equation}
m_{a}(0,r)=m_{a}(0,a)+\Delta E_{a}(a)\,.  \label{16}
\end{equation}

At the end of the fall, according to special relativity applied
locally at $a$,

\noindent 
\begin{equation}
m_{a}(V,a)=\gamma m_{a}(0,a)=m_{a}(0,a)+\Delta E_{a}(a)\,.  \label{17}
\end{equation}

By comparing (\ref{16}) and (\ref{17}),

\noindent 
\begin{equation}
m_{a}(0,r)=m_{a}(V,a)\,.  \label{18}
\end{equation}

This means that:
\begin{itemize}
  \item {\it During a free fall, the mass-energy of a body,
with respect to any SI observer, remains constant and equal to its initial
rest mass with respect to the same observer.} (Conservation of NL
mass-energy with respect to an observer in a fixed position in the field).
\end{itemize}

On the other hand, from (\ref{17}) and (\ref{18}), the net energy released 
during the stop is:

\noindent 
\begin{equation}
\Delta E_{a}(a)=m_{a}(V,a)-m_{a}(0,a)=m_{a}(0,r)- m_{a}(0,a)\,.  \label{19}
\end{equation}

\begin{itemize}
  \item  {\it The net energy released during a free fall comes
``not'' from the G field. It is a small fraction of the original 
mass-energy of the body}.

  \item  There is not a true exchange of energy between the field
and the body.
\end{itemize}

\subsection{Demonstrations from gedanken-experiments}

   \subsubsection{Deduction from matter annihilation during a free fall}

\noindent
Assume an observer at rest at a distance $a$ from a neutron star, 
relatively far from it. He makes a global mass-energy balance of the 
system after the free fall of an electron-positron pair. Statistically, 
the electron pair may decay in any arbitrary radius $r$. For simplicity, 
assume that it decays into two gamma photons travelling symmetrically 
with respect to the vertical.

According to global mass-energy conservation in the system, the net 
energy crossing the imaginary sphere of radius $a$ is independent on 
the actual radius in which annihilation actually occurs. Thus either 
for the annihilation occurring at an arbitrary radius $r$, or for the 
one occurring at $a$, the net energy going away from the
system must be the same and equal to:

\noindent 
\begin{equation}
m_{a}(V,r)= 2h\nu_{a}(r)= 2h\nu_{a}(a)= m_{a}(0,a)\,.  \label{20}
\end{equation}

The energy released by NL annihilation occurring at $r$, with
respect to the observer A, is stated in the first and second member. This
one must be is equal to the one released by annihilation at the observer's
radius, which is stated in the 3rd and 4th member.

Notice that the NL frequency conservation of the photons is
obvious from the 2nd and 3rd member. The NL mass-energy conservation of the
body is obvious from the first and the last member.

Then from these gedanken-experiments is concluded that {\it %
both, the NL mass-energies of the bodies and the NL energies of the photons,
with respect to an observer in a fixed potential, remain constant during
their respective trips in the field.}

\subsubsection{Deduction from radioactive decay during a free fall}

\noindent
The observer in the previous experiment may also
study the free fall of a radioactive atom that generates a gamma photon in
any arbitrary radius of a central field.

According to the EP, the energy of a photon is a constant
fraction of the mass of the atom that emits it. This fraction is independent
both on the observer's position and on the radial position in which the atom
decays. Then, for a decay at $a$ and for the decay at $r$, this common ratio
is:

\noindent 
\begin{equation}
\frac{\Delta E}{m}= \frac{h\nu_{a}(a)} {m_{a}(0,a)}= \frac{h\nu_{a}(r)}{%
m_{a}(V.r)}= Constant\,.  \label{21}
\end{equation}

From Eq.~(\ref{13}), the numerators in the Eq.~(\ref{21}) must
be the same. Consequently, the denominators of such equation must also have
the same values. Then it is verified, again, that ``during the free fall,
the mass-energy of an atom, with respect to a SI observer, does not
increase. It remains invariable''.

\section{Can the Distances Increase Due to Universe Expansion?}

\noindent
In the current theories on universe expansion it is tacitly assumed, without 
fair reasons, that {\it some arbitrary kinds of bodies do not expand in the 
same proportion as some other parts of the universe}. This is equivalent
to assume that particles are invariable after a change of G potential, which 
is in clear contradiction with the above transformations.

From Eqs.~(\ref{7})and (\ref{11}), when a measuring rod gets away from 
another the G field source, the body is expanded in the proportion:

\noindent 
\begin{equation}
\frac{dL_{a}(0,r)}{L_{a}(0,r)}= \frac{d\lambda _{a}(0,r)}{\lambda _{a}(0,r)}%
= d\phi (r)={\frac{GM}{r}}\frac{dr}{r}\,.  \label{21b}
\end{equation}

Then, there should be a phenomenon of G expansion of matter
associated to the general increase of G potential produced by the increase
of the distances all of the bodies of the universe. Such expansion is not
negligible because the contribution of shells of matter of increasing radius
also increases with their radius. On the other hand, the contribution of 
each shell must be corrected for cosmological redshift (or Doppler shift) 
which is proportional to the trip of light, i.e., proportional to 
$exp-[r/R]$, in which $R$ is the Hubble radius. Thus, after plain integration, 
it is simple to find that {\it every particle of it expands in just the same 
proportion as any other distance of the 
universe}\cite{Vera96a}$^,$\cite{Vera97}. This means that the ratio between 
distances must remain unchanged, indefinitely.

The same conclusion comes out more straightforwardly by emulating every 
particle of the universe by a quantum in stationary state.

It is important to remember here that, according to current interference and 
diffraction experiments, the position of a photon is fixed by the result of 
constructive interference of more elemental kinds of ``wavelets'' that travel 
with the speed of light. In principle the isolated wavelets have no energy. 
They are not destroyed during the interference phenomenon. 
Then the stationary photon of each particle model must also be the
result of a constructive interference of wavelets travelling in opposite
directions. Such wavelets, after constructive interference in the model,
must travel rather indefinitely in the space. Then the empty space must be
crossed with a high flux of wavelets with different frequencies and {\it
random phases}. They come from all of the particles of the universe. Thus
the space turns out to be crossed by a high flux of ``wavelets'' coming back
and forth from all of the particles of universe. They must interfere
constructively only in the sites of particles and free radiation. In the
rest of the space, they must interfere with ``random phases''.

According to ``wavelet-continuity'', or Doppler effect, a
uniform universe expansion would enlarge every length and wavelength,
without exception, in just the same proportion. Then the universe expansion
would not change the relative numbers of waves between and within the
particles, i.e., it would not change the ratio between distances. Any
measuring rod would also expand in just the same proportion, so that the
average distances and the average density of the universe should also 
remain invariable throughout the time. Then, from a relative viewpoint, 
the ``average'' universe must look like it was static forever. 

On the other hand, the universe expansion is not the single alternative
for explaining the cosmological redshift. Any kind redshift of photons 
proportional to the distances of propagation would account for it 
provided that the statistical fluctuations do not widen the spectral 
lines. Thus the possibilities for interactions of the photon's wavelets 
with the uniform wavelet background of the rest of the universe 
cannot be arbitrarily ruled out. If that were true, according to
interference laws, such interaction would transfer energy from the 
photon to the rest of the universe, with almost no statistical 
fluctuation.

Consequently, whatever is the ultimate nature of the Hubble redshift, 
the universe age must be many orders of magnitude higher than the 
current estimations, may be infinity.

\subsection{The ``linear'' kind of black hole fixed by
the experimental facts}

\noindent
The G transformations fixed by the experimental facts are strictly linear 
ones, i.e., without singularity.. Thus the properties of a ``linear'' 
black hole (LBH) turn out to be radically different from the ones
of the conventional black hole. For example, {\it the local physical laws
must still hold within and around a neutron star with $2GM>>r$}. 
So that ``the Fermi exclusion principle would prevent 
collapse''\cite{Ahluwalia 99}. 
On the other hand, from Eq.~(\ref{10}) it is simple to find that 
the gradient of the NL refraction
index of the space around such macronucleus produces a critical reflection
towards the inside of field thus preventing an appreciable escape of any
kind of radiation\cite{Vera81b}. 

Thus the LBH must be like a giant macronucleus that ``absorbs''
energy from the rest of the universe until it can explode. The final
expansion can only occur after a very long period in which the average
mass-energy per nucleon, with respect to an external observer, becomes
higher than the one of H in free state.

During the LBH expansion, the nucleons must decay into new H,
rather free of heavy metals. The cluster of dead star remnants, normally
travelling around a LBH, must condense such gas. This process should
regenerate star clusters and galaxies with the maximum proportions of clean
gas and angular momentum with random orientations. Such properties are
consistent with those of globular clusters and some elliptical galaxies.
Thus they are clear testimonies of relatively recent explosions of single or
chain of LBH explosions that have occurred everywhere in the 
universe\cite{Vera81b}$^,$\cite{Vera97}.

\section{The new cosmological scenario fixed by the experiments}

\noindent Due to the linear properties of the black holes, and
to the rather infinite age of the universe, the new cosmological scenario is
necessarily different from the conventional one.

The atoms, throughout star evolution, must be evolving, rather
indefinitely, in nearly closed cycles between the states of gas and the new
kind of ``linear'' black hole, and vice versa. The last ones, after
absorbing radiation from the rest of the universe, must explode regenerating
gas with maximum densities of randomly oriented angular momentum. Such
explosions, and chains of them, must fix the initial 
stages of the evolution cycles of star
clusters and galaxies. Inevitably, the last ones must also evolve in rather
closed cycles between luminous and non luminous stages cooled down by LBHs.
They may be called ``black galaxies''.

Statistically, each stage of a galactic cycle should be present
in the sky in a proportion according to its respective evolution period.
Since the black period of a galaxy must be of higher orders of magnitude
that the luminous one, then most of the galaxies of the universe must be in
the state of ``black galaxy'' cooled down by the LBHs generated after
evolution of the most massive stars. They must be absorbing energy from the
rest of the universe. They must account for the missing mass problem in the
intergalactic space and for the large number of phenomena observed in such
space.

Since the black galaxies must cooled down by LBHs and by the
rest of the universe, then the dead bodies orbiting around must 
have very low average temperatures.
Since they are in very low G potentials, they must emit blackbody radiation
with some appreciable G redshift. Additionally, it is simple to find that
most of such radiation must come from distance ranges of the order of
magnitude of about a Hubble radius i.e., it must be strongly redshifted by
the cosmological redshift\cite{Vera96a}$^,$\cite{Vera97}. Then such radiation 
must correspond with the low temperature CMBR observed in the universe.

Due to the strict linearity of the G transformations, the black galaxies 
would not dissipate energy in the form of G waves. Thus they would not 
collapse during the long periods in which they recover energy from the 
rest of the universe.

\section{Discussion}

\noindent
The phenomenological reasons for the above results
can be studied by using the particle model cited above. This one is made up of 
one or more photons in stationary state between any two parts (or mirrors) of
the same system. According to the EP, such model must obey the same inertial
and gravitational laws as any other part of the same system.

The basic parameters of a local particle model at rest in the
field are its frequency, $\nu_{a}(0,a)$, and its wavelength, $\lambda_{a}(0,a)$.
Its rest mass-energy is $m_{a}(0,a)=nh\nu_{a}(0,a)$.

From Eq.~(\ref{10}), the NL parameters of such particle at rest with
respect to a SI observer, are position dependent. They can also be expressed 
as functions on the square root of the
NL speed of light of the space in which it is located\footnote{
Vera has verified that that the theoretical properties of the particle model
are consistent with special relativity, quantum mechanics and with the
conventional 
G tests\cite{Vera81b}$^,$\cite{Vera97}.}. Then they can be used to
explain the above results.

\subsection{Why the G field does not exchange energy with the particle model?}

\noindent
According to the nature of the particle model, its G field can only depend 
on long range properties of photons, which are
currently described in terms of the ``wavelets'' used in optical physics.
From the form of the G potential it is inferred that the average properties
of the space around a particle model can only depend on the perturbation
rate produced by all of the wavelets with ``random phases'' that are
actually crossing it. Each wavelet contribution would be proportional to the
product of its frequency and to its amplitude. According to this, and to the
Huygen's principle, the probability for the existence of real energy in the
empty space, away from the model, is proportional to the net wavelet
amplitude, which in this case is zero. Statistically, this means that {\it 
the probability to find energy in the empty space of a G field, away from a
stable particle, is null}. This would be a fundamental difference with
electric fields.

Then ``the G field cannot exchange energy with bodies or photons 
because it has no energy''.

This is also clear from Eq. (\ref{10}). According to it, 
the G field is a gradient of the NL refraction index of 
the space. The same as in any refraction phenomenon, and 
according to wavelet continuity, the frequency photons 
travelling throughout a perfect dielectric does not change 
Then there is no energy exchange between the field and the 
particle model. However there is a momentum exchange that 
occurs during the internal trips of
the waves\cite{Vera81b}$^,$\cite{Vera97}.

\subsection{The origin of conventional errors}

\noindent
The original error is inherited from classical
physics in which matter was assumed to be absolutely invariable. Later on,
after special relativity, the EP was erroneously interpreted as a statement
on the absolute invariability of the parameters of bodies after a change of
G potential. Thus all of the observers arbitrarily assigned the same
numerical values to their standard clocks and standard atoms.

But such principle admits two alternative hypotheses:

I) The conventional one, in which ``all of the bodies are
absolutely invariable after a change of G potential'',

II) The opposed one, in which ``all of the bodies change in
just the same form and in an identical proportion''.

According to the conventional alternative, the parameters of a
NL body are assumed to be functions on just its velocity. They are tacitly
assumed to not depend on its position in the field. According to the second
alternative, the parameters of a body, with respect to the observer, depend
on the positions of the body and of the observer.

In a ``free fall'', for example, the conventional mass-energy
balance may have the form:

\noindent 
\begin{equation}
m(V)=m(0)+\Delta E\,.  \label{22}
\end{equation}

In which $m(V)$ is the (relativistic) mass measured by an 
observer $A$ at the end of the fall, $m(0)$ is the initial rest 
mass ``measured by another observer $B$'', and $\Delta E$ is 
the energy released during the stop and measured by $A$.
This equation currently makes believe in that the relativistic 
mass of the body has increased, during fall, due to some presumed 
energy $\Delta E$ given up by the G field to the body.

The lack of a strict homogeneity of this equation stands out
when the distances of the object ($b$) and of the observer ($a$), with respect
to the central field source, are explicitly stated:

\noindent 
\begin{equation}
m_{a}(V,a)=m_{b}(0,b)+\Delta E_{a}\,.  \label{23}
\end{equation}

The different subscript of the mass in the second member puts
on evidence that this equation is a mixture of quantities referred to two
different reference standards located at the distances $a$ and $b$ from the
G field source, respectively. According to the experimental evidences, and
from Eq. (\ref{9}), the reference standard of the observer at $b$ is
physically different with respect to the one of the observer at $a$. Then,
strictly, this equation has no well-defined physical meaning.

It may be argued that the value of $m_{b}(0,b)$ can be replaced
by $m_{a}(0,a)$, because they have the same numerical values. Effectively,
such replacement gives a more strictly homogeneous relation:

\noindent 
\begin{equation}
m_{a}(V,a)=m_{a}(0,a)+\Delta E_{a}\,.  \label{24}
\end{equation}

But the physical meaning of (\ref{24}) is radically different 
from above one. This is a strictly ``local'' mass-energy balance
``during the stop at $a$. From just Eq.~(\ref{24}), it is not possible to
find whether the kinetic energy comes from the body of from the G field.
Thus, to find where such energy comes from, some other better defined
experiment, such as GTD, or the above transformations, must be used.

Then it is obvious that the hypothesis on the G field energy,
used by Einstein to postulate his field equation\cite{Einstein}, is a direct
consequence of a
tacit postulate on the absolute invariability of matter after a change of G
potential. Indeed, he assumed that ``the G field transfers energy and
momentum to the matter in that it exerts forces upon it and give it give it
energy''. The reason given by him is ambiguous because to exert forces is
not synonymous of giving up energy. A static road, for example, exerts
forces on a runner that is accelerating by himself. Such forces do not give
up energy to the runner. The runner puts on the energy for his acceleration,
the same as the body in the case of a free fall.

Later on the equations used for conventional tests for G
theories were based on position independent parameters, i.e., they were
based on the same tacit postulate on the invariability of matter that was
used in GR\footnote{For example, in the experiments on GRS of
photons\cite{Pound} it is tacitly assumed that ``the
clocks in different distances from the earth run with the same frequency
with respect to each other", which is not consistent with the true GTD
experiments.}. Then, in one way or another, ``the G tests have been partial
to GR because they depend on the same kind of error. On the other hand they
have helped to reject other theories that were free of such error. This
seems to be one of the reasons for which this error has prevailed for about
one century.

\section{Conclusions}

\noindent

According to the experimental facts, observers in different G
potentials have reference standards that are not physically the same with
respect to each other. Thus the current relationships between quantities
measured in different potentials are inhomogeneous and without well-defined
physical meanings. When the observers change of position in the field, they
cannot detect such changes from just local measurements because all of the
parts of their measuring systems change in the same way and in same
proportion after any common change of position in the field, i.e., every
ratio remains unchanged

Consequently, to relate quantities measured by observers in
different positions of a static G field, they must be previously transformed
to some common physical unit system based on a SI reference standard in some
fixed and well-defined position (or distance) with respect to the G field
source.

According the new scenario fixed by the experimental facts:

\begin{enumerate}
\item  The bodies are not invariable after a change of position
with respect to the G field source and with respect to the observer. Thus
the parameters of NL bodies with respect to an observer are ``position
dependent''.

\item  The basic parameters of any well-defined part of a
measuring system change in the same proportion after a common change of
position with respect to the G field source. Such parameters are ``position
dependent''.

\item  The NL speed of a photon in a G field, with respect to
an observer in a fixed position, is not constant. It is ``position
dependent''.

\item  The parameters of the NL bodies depend on the square
root of the NL speed of light in such position.

\item  Photons and bodies do not exchange energy with the G field. They
are just ``refracted'' by a gradient of the NL refraction index of the space,
i.e., they exchange just momentum. Their NL frequencies and 
mass-energies, with respect to SI
observers, remain constant (NL frequency conservation)\footnote{%
This is, anyway, a direct consequence of wave continuity of light travelling
in strictly conservative fields, which is a fundamental property of light.
This conservative property has been used before for proving that matter is
not invariable after a change of G potential\cite{Vera81b}}.

\item  During a free fall, the mass-energy of a body, with
respect to a fixed observer, does not increase. It remains constant (NL
mass-energy conservation.).

\item  During G work, the G field does not give up energy to
the bodies. The field puts on just the momentum that can release the body's
mass-energy. The energy coming from G work is a small fraction of the
original mass-energy of the body that, most probably, is confined as
stationary quanta.

\item  Particles can be emulated by particle models made up of photons 
in stationary state.

\item  According to the nature of the particle model, the G field cannot 
exchange energy with photons and bodies because it has no energy.

\end{enumerate}

In cosmology, also, the conventional properties of the black
holes and of the universe strongly depend on the tacit postulate on the
invariability of matter. 

According the linear relationships fixed by the EP, the LBHs absorb radiation 
from the rest of the universe up to a critical point in which they can 
explode. Thus they return the nucleons to the state of H gas. 

If the universe where expanding, the increase of G potential must produce a 
G expansion of every particle and distance between them, in just the same 
proportion, so that the average distance ratios must remain unchanged, 
in contradiction with the original hypothesis. 

Thus, most probably the cosmological redshift is due to some unknown 
interaction of the photon's wavelets with the wavelets of the
rest of the universe. This one would not produce appreciable line 
widening. In any case, the true age and lifetime 
of the universe must be rather unlimited.

However the EP and the experimental facts fix well-defined properties of 
the LBHs and of the ``actual'' universe, which in turn fix the global way
according to which the celestial bodies must evolve throughout the time.

Statistically, atomic matter must evolve in rather
closed cycles between the states of gas and LBH, and vice versa. The
explosions of LBHs fix the initial periods of the evolution cycles of
luminous star clusters and galaxies. Thus the last ones must also evolve,
indefinitely, in rather closed cycles between luminous states and black
states cooled down by LBHs. Something similar holds for clusters.

Thus the new cosmological scenario fixed by the true G tests
has fundamental differences with respect to the conventional 
one\cite{Vera81b}$^,$\cite{Vera97}. For example:

\begin{itemize}
\item  {\it Only a very small fraction of the universe must be
in luminous states}. Most of it must be in the state of compact black 
galaxy. They must be the end products of the normal evolution of 
previous luminous galaxies. They must be
cooled down by the LBHs coming from natural evolution of their most massive
stars and star clusters. They must account for all of them: the low 
temperature and the missing mass of the intergalactic space, and for 
the phenomena occurring in such region, like gamma bursts, cosmic 
radiation and iron spectrum. Thus {\it the
low temperature CMBR is not a cosmic relic but just a fair testimony of the
high proportion of galaxies cooled down by LBHs}.

\item  {\it The stars would be not formed from a real primeval
gas, after long periods in a seedless space}. They must come from
relatively fast condensation of gas, recently ejected from LBHs or massive
stars, over other bodies that were travelling around\footnote{
The older seed bodies, like planets and dead stars, must work in a way 
similar way to the ``getters'' used
to remove the gas in a vaccum tube.}. Thus the most recently formed star
clusters and galaxies can be recognized by the highest proportions of the
randomly oriented angular momentum, and by the lowest contamination with
metals. They are clear evidences in that such gas comes from recent
explosions of LBHs\cite{Vera01}.

\item  In a matter cycle, the condensation of gas up to LBH state 
produces {\it a G energy yield of a higher order of magnitude than that of 
nuclear fusion of H}. Such additional energy must be able to regenerate
H, from He, after nuclear stripping, and to generate cosmic radiation around 
neutron stars. It must also account for the higher energy yields and the 
higher luminous lifetimes of some regions of the 
galaxies\cite{Vera93}$^,$\cite{Vera97}$^,$\cite{Vera01}.

\item  The entropy of the universe must not increase with the
time. The LBHs must be cleaning up the space from all kinds of quanta,
regardless on their low energies. Such energy must be filling up energy
levels of neutrons in the LBHs. Such energy must appear, again, after their
explosions, during star evolution.
\end{itemize}

Globally, the EP and the true G tests also fix, independently on any G 
theory, the invariability of the average parameters of the universe, 
like the mass-energy, density and entropy. 

Curiously, the disagreements of the conventional scenarios with the 
experimental facts come, ultimately, from the tacit postulate on the 
absolute invariability of the atoms and clocks after a change of position 
with respect to the G field sources. 

\nonumsection{References}

\end{document}